\documentclass[twocolumn,showpacs,preprintnumbers,amsmath,amssymb]{revtex4}
\usepackage{graphicx}
\usepackage{dcolumn}
\usepackage{bm}

\usepackage{ifthen}
\usepackage{epsfig}
\usepackage{rotating}
\usepackage{hyperref}

\newcommand {\color}{color}

\begin{document}
\newcommand{\TOPAZref}    {Montagna:1993py, Montagna:1993ai, Montagna:1996ja, Montagna:1998kp}
\newcommand{\ZFITTERref}  {Bardin:1989di, Bardin:1990tq, Bardin:1991fu, Bardin:1991de, Bardin:1992jc, Bardin:1999yd}
\newcommand{\antibar}[1]{\ensuremath{\mathrm{#1\overline{#1}}}}
\newcommand{\Zpole}{\mbox{\Zzero-pole}}
\newcommand{\MH}{m_{\mathrm{H}}}
\newcommand{\Mt}{m_{\mathrm{t}}}
\newcommand{\MZ} {\ensuremath{m_{\mathrm{Z}}}}
\newcommand{\Zzero}{\ensuremath{\mathrm{Z}}}
\newcommand{\swsq}{\sin^2\thw}
\newcommand{\MW} {\ensuremath{m_{\mathrm{W}}}}
\newcommand{\GF}{G_{\mathrm{F}}}
\newcommand{\cwsq}{\cos^2\thw}
\newcommand{\rhoo}{\rho_0}
\newcommand{\thw}{\theta_{\mathrm{W}}}
\newcommand{\alqed}{\alpha(\MZ^2)}
\newcommand{\ee}{\ensuremath{\mathrm{e}^+\mathrm{e}^-}}
\newcommand{\ff}{\antibar{f}}
\newcommand{\bb}{\antibar{b}}
\newcommand{\mumu}{\ensuremath{\mu^+\mu^-}}
\newcommand{\tautau}{\ensuremath{\tau^+\tau^-}}
\newcommand{\dalhad}{\Delta\alpha^{(5)}_{\mathrm{had}}(\MZ^2)}
\newcommand{\Drw}{\Delta{r}_{\mathrm{w}}}
\newcommand{\rhof}{\rho_{\mathrm{f}}}
\newcommand{\kappaf}{\kappa_{\mathrm{f}}}
\newcommand{\rhose}{\rho_{\mathrm{se}}}
\newcommand{\kappase}{\kappa_{\mathrm{se}}}
\newcommand{\swsqefff}{\sin^2\thwefff}
\newcommand{\gaf}{g_{\mathrm{Af}}}
\newcommand{\gvf}{g_{\mathrm{Vf}}}
\newcommand{\gal}{g_{\mathrm{A\ell}}}
\newcommand{\gvl}{g_{\mathrm{V\ell}}}
\newcommand{\thwefff}{\theta_{\mathrm{eff}}^{{\rm f}}}
\newcommand{\Gff}{\Gamma_{\mathrm{f\bar{f}}}}
\newcommand{\Gll}{\Gamma_{\ell\ell}}
\newcommand{\swsqeffl}{\sin^2\theta_{\mathrm{eff}}^{\mathrm{lept}}}
\newcommand{\Rbz}{\ensuremath{R_{\mathrm{b}}^0}}
\newcommand{\Rcz}{\ensuremath{R_{\mathrm{c}}^0}}
\newcommand{\Gbb}{\ensuremath{\Gamma_{\bb}}}
\newcommand{\Ghad}{\Gamma_{\mathrm{had}}}
\newcommand{\Ginv} {\Gamma_{\mathrm{inv}}}
\newcommand{\GZ}{\Gamma_{\mathrm{Z}}}
\newcommand{\shad}{\sigma_{\mathrm{h}}^{0}}
\newcommand{\Rl}{R^0_{\ell}}
\newcommand{\Afbzl}{\ensuremath{A^{0,\,\ell}_{\mathrm{FB}}}}
\newcommand{\cAl}{\ensuremath{{\cal A}_{\ell}}}
\newcommand{\avQfb}{{\langle Q_{\mathrm{FB}} \rangle}}
\newcommand{\Afbzb}{\ensuremath{A^{0,\,{\mathrm{b}}}_{\mathrm{FB}}}}
\newcommand{\Afbzc}{\ensuremath{A^{0,\,{\mathrm{c}}}_{\mathrm{FB}}}}
\newcommand{\cAb}{\ensuremath{{\cal A}_{\mathrm{b}}}}
\newcommand{\cAc}{\ensuremath{{\cal A}_{\mathrm{c}}}}
\newcommand{\alfas}{\alpha_{\mathrm{S}}}
\newcommand{\ALR}{\ensuremath{A_{\mathrm{LR}}}}
\newcommand{\LOGMH}{\log_{10}(\MH/\GeV)}
\newcommand{\GeV}{\mathrm{GeV}}
\newcommand{\MeV}{\mathrm{MeV}}
\newcommand{\pp}{\antibar{p}}
\newcommand{\cgaf}{{\cal G}_{\mathrm{Af}}}
\newcommand{\cgvf}{{\cal G}_{\mathrm{Vf}}}
\preprint{hep-ex/0402039}
\preprint{UMD PP-04 018}
\preprint{UT-ICEPP 04-01}

\title { The Character of $\Zpole$ Data Constraints on\\
 Standard Model Parameters\\ 
}

\author{ T. Kawamoto}
 \email{tatsuo.kawamoto@cern.ch}
\affiliation{%
International Centre for Elementary Particle Physics,\\
University of Tokyo, Tokyo 113-0033, Japan}

\author{R.G. Kellogg}
 \email{richard.kellogg@cern.ch}
\affiliation{
Department of Physics, University of Maryland,\\
College Park, MD 20742, USA}

\date{\today}

\begin{abstract}
  
Despite the impressive precision of the $\Zpole$ measurements made
with the e+ e- colliders at CERN (LEP) and SLAC (SLC),
the allowed region for the principle
Standard Model parameters responsible for radiative corrections
($\MH$, $\Mt$, and $\alpha(\MZ^2)$) is still
large enough to encompass significant non-linearities.
The nature of the experimental constraints therefore depends in an interesting
way on the ``accidental'' relationships among the various measurements.
In particular, the fact that the $\Zpole$ measurements favor values
of $\MH$ excluded by direct searches leads us to examine the
effects of external Higgsstrahlung, a process ignored by the
usual precision electroweak calculations.
\end{abstract}

\maketitle

\section{Introduction}

The precise measurements of $\Zzero$ resonance properties made by the four
collaborations at the CERN $\ee$ collider LEP and by the SLAC Large Detector
(SLD) at the SLC
have confirmed all relevant predictions of the Standard Model,
and established a strong experimental basis for the mechanism of
symmetry breaking in the electroweak sector.
The two fundamental equations of electroweak unification,
\begin{eqnarray}
  \label{eq:GF}
  \swsq &=& \frac{\pi\alpha}{\sqrt{2}\MW^2\GF},\\
  \label{eq:rho}
  \cwsq &=& \frac{\MW^2}{\MZ^2\rhoo},
\end{eqnarray}
establish the relations between the strengths of the electromagnetic and weak
couplings, and the mass ratio of the neutral and charged heavy vector bosons.
Here $\GF$ is the Fermi constant determined in muon decay, $\alpha$
is the electromagnetic fine-structure constant, $\MW$ and $\MZ$ are the W and
$\Zzero$ boson masses, and $\thw$ is the electroweak mixing angle.
The $\rhoo$ parameter\cite{Ross:1975fq} is determined by the Higgs structure
of the
theory; in the Minimal Standard Model which we consider here,
$\rhoo$ is unity.

$\GF$, $\MZ$, and $\alqed$ are known with relative precisions of
(1, 2 and 40)$\times10^{-5}$.
This allows the radiative corrections to Equations~\ref{eq:GF}
and~\ref{eq:rho} to be investigated in considerable detail.
These radiative corrections 
depend principally on the masses of the top quark, $\Mt$,
and the Higgs boson, $\MH$. 
One of the great strengths of
the $\Zpole$ data is that the measurements of the partial widths and charge
asymmetries allow the effects of the Higgs boson and the top quark to be
separated to a remarkable extent.
The character of this separation is the primary topic of this paper.

Since the top quark has been observed directly, the agreement between
the directly measured mass and its indirect measurement derived through the
radiative corrections provides compelling evidence that the theoretical 
and experimental understanding of the $\Zpole$ measurements rests upon firm
ground.
The much smaller effects due to $\MH$ then provide essentially the only
experimental knowledge we have concerning this elusive particle, apart
from the fact that it has so far escaped direct observation.

The broad range of $\MH$ still consistent with the measurements provides 
sufficient room for non-linear effects to become important.
Such non-linearities mean that the gaussian error hypothesis, which is implicit
in the $\chi^2$ fits used to determine the error contours of the
Standard Model parameters, approximates the true errors
only imperfectly. The actual errors and even the shape of the measurement
constraints depend significantly on the working point of the fit in
the multi-dimensional space describing the range accessible to the
parameters.
The character of the $\Mt$, $\MH$ separation therefore depends in an
interesting manner on the ``accidental'' relations between the various
measurements.
Variations on the scale of the expected measurement uncertainties can,
in some cases, lead to disproportionate shifts in the error contours.
In other cases, the specific manner in which a measurement happens to lie
within its band of uncertainty results in unanticipated stability.
Some understanding of these subtle effects is necessary for a full
appreciation of the measurements.

Since we are interested in describing how our knowledge is augmented by
each individual measurement as well as sets of measurements in combination,
we are led naturally to consider hypothetical situations which explicitly
violate constraints lying beyond the set of measurements being
considered at any particular moment.
Chief among the constraints we often ignore is the experimental
lower limit on $\MH$ provided by the  direct searches~\cite{LEPSMHIGGS}.
A related complication in the analysis arises
from what may be an historical accident. Due to the early and continued
non-observation of the Higgs boson at LEP, the two most precise electroweak
programs, ZFITTER~\cite{\ZFITTERref} and TOPAZ0~\cite{\TOPAZref}, do not
include the effects of external Higgs boson emission from the $\Zzero$
propagator.
Since the $\Zpole$ measurements in fact favor a low-mass Higgs boson,
we are led to consider the effects of such Higgsstrahlung in order to
provide a self-consistent picture of what these measurements alone
tell us about the Standard Model.

\section{Radiative Corrections}
\label{sec:rad-cor}

The largest radiative correction affecting the basic electroweak relations
presented in the previous section is due to the running of the
electromagnetic coupling constant in Equation~\ref{eq:GF}.
This running is due to the presence of fermion loops in the photon
propagator, and is usually parametrized as:
\begin{equation}
  \label{eq:alpharun}
  \alpha(s) = \frac{\alpha(0)}{1-\Delta\alpha(s)}.
\end{equation}
Near the $\Zpole$, leptons, the top quark, and the five light quarks contribute to
$\Delta\alpha(s)$:
\begin{equation}
  \label{eq:Dalpha}
  \Delta\alpha(s) = \Delta\alpha_{\mathrm{e\mu\tau}}(s)
                   +\Delta\alpha_{\mathrm{top     }}(s)
                   +\Delta\alpha_{\mathrm{had     }}^{(5)}(s).
\end{equation}
The $\Zpole$ data itself gives no useful experimental constraints on $\Delta\alpha(s)$.
The first two terms of Equation~\ref{eq:Dalpha} can be precisely calculated,
but $\Delta\alpha_{\mathrm{had}}^{(5)}(s)$ is best determined by analyzing
the measured rate of $\ee$ annihilation to hadrons using a dispersion
relation~\cite{bib-BP01}, where low-energy data dominates the resulting
uncertainty.
At LEP/SLC energies, $\alpha$ is increased from the Thompson limit of
$1/137.036$ to $1/128.945$, corresponding to
$\dalhad = 0.02761\pm0.00036$~\cite{bib-BP01}.

Combining Equations~\ref{eq:GF} and~\ref{eq:rho}, and including radiative
corrections yields:
\begin{eqnarray}
  \label{eq:GFmod}
 \cwsq\swsq & = &
\frac{\pi\alpha(0)}{\sqrt{2}\MZ^2\GF}\frac{1}{1-(\Delta\alpha  + \Drw)} ,
\end{eqnarray}
where $\Drw$ represents further weak corrections.
Equation~\ref{eq:GFmod} directly links $\Delta\alpha$ and $\Drw$ to the
value of $\thw$, and hence  to $\MW$ via Equation~\ref{eq:rho}.

Radiative corrections also affect the couplings of the $\Zzero$ to fermions.
The bulk of these corrections\cite{Veltman:1977kh}
can be absorbed into an over-all scale factor for the couplings,
$\rhof$, and a scale factor, $\kappaf$, for the
 electroweak mixing angle, resulting in
``effective'' quantities:
\begin{eqnarray}
 \rhof &=& 1  + \Delta\rhose + \Delta\rhof\label{eq:rhoeff}\\ 
 \kappaf &=& 1  + \Delta\kappase + \Delta\kappaf\label{eq:kappaeff}\\ 
  \label{eq:sin}
   \swsqefff &=& \kappaf\swsq\\
  \label{eq:gveff}
  \gvf &=& \sqrt{\rhof}(T_f^3 - 2Q_f\swsqefff) \\
  \label{eq:gaeff}
  \gaf &=& \sqrt{\rhof}T_f^3.
\end{eqnarray}
Here $T_f^3$ is the third component of weak isospin, $Q_f$ is the charge,
$\thwefff$ is the effective electroweak mixing angle, and $\gvf$ and $\gaf$
are the effective vector and axial-vector couplings for the
fermion species f.
The quantities $\Delta\rhose$ and $\Delta\kappase$ are universal corrections
arising from the
propagator self-energies, while $\Delta\rhof$ and $\Delta\kappaf$ are
flavor-specific vertex corrections.
The effective couplings $\gvf$ and $\gaf$ are purely real, and describe
that part of the $\Zzero$ interaction which can be treated as a
resonance in the Improved Born Approximation. The remaining,
complex parts, termed ``remnants'', generate effects which are small compared to the
experimental errors, but are nevertheless respected when determining $\gvf$ and $\gaf$
from the data (see, for example~\cite{Abbiendi:2000hu}).

The $\Zzero$ partial width to each fermion species, $\Gff$, is proportional to
$\rhof$, while $\swsqeffl$, measured through the asymmetries, is proportional to $\kappaf$.
Calculations at one-loop order illustrate the essential $\Mt$ and $\MH$
dependencies\cite{Burgers:LEP1YR89VOL1}, which are quadratic in $\Mt$ and
logarithmic in $\MH$.
The dependence of $\Delta\kappase$ remains the same over the entire range of $\MH$:
\begin{eqnarray}
   \label{eq:deltakappa}
 \Delta\kappase  = &&
     \frac{3\GF\MW^2}{8\sqrt{2}\pi^2}\frac{\Mt^2}{\MW^2}\frac{\cwsq}{\swsq} -
     \nonumber \\ 
     &&\frac{10}{3}\frac{\GF\MW^2}{8\sqrt{2}\pi^2}\left(\ln\frac{\MH^2}{\MW^2}
-\frac{5}{6}\right) + \cdots .
\end{eqnarray}
For $\Delta\rhose$, the sign of the dependence on $ln(\MH)$ is negative for
$\MH \gg \MW$:
\begin{eqnarray}
   \label{eq:deltarho}
    \Delta\rhose  = &&
 \frac{3\GF\MW^2}{8\sqrt{2}\pi^2}\frac{\Mt^2}{\MW^2} -
     \nonumber \\ 
    &&\frac{3\GF\MW^2}{8\sqrt{2}\pi^2}
    \frac{\swsq}{\cwsq}\left(\ln\frac{\MH^2}{\MW^2} -\frac{5}{6}\right) +
\cdots  
\end{eqnarray}
and positive for $\MH \ll \MW$:
\begin{eqnarray}
   \label{eq:deltarhomhsmall}
    \Delta\rhose  = &&
\frac{3\GF\MW^2}{8\sqrt{2}\pi^2}\frac{\Mt^2}{\MW^2} +
     \nonumber \\ 
&&\frac{\GF\MZ^2}{4\sqrt{2}\pi^2}\ln\frac{\MH^2}{\MZ^2} -  
\frac{7\GF\MH\MZ}{8\sqrt{2}\pi} \cdots . 
\end{eqnarray}

Equation~\ref{eq:sin} shows that $\swsqeffl$ receives radiative corrections from
both $\Drw$, through $\swsq$, and from $\Delta\kappase$ directly.
These corrections act along the same axis in the $\Mt\,vs\,\ln(\MH)$ plane, but
in opposite senses due to the near equality,
in single-loop approximation:
\begin{eqnarray}
  \label{eq:kappadrw}
  \Delta\kappase   &=& -\Drw + \cdots.
\end{eqnarray}
In the $\Mt\,vs\,\ln(\MH)$ plane lines of constant $\Delta\kappase$ and $\Drw$,
and hence $\MW$ and $\swsqeffl$ are therefore all approximately parallel.
When moving along such lines, then, changes in the radiative effects of the
top quark and the Higgs boson are seen to cancel.
It is only through violations of this approximation that measurements of $\MW$ and
$\swsqeffl$ have any power to separate the effects of $\Mt$ and $\MH$.

The fact that $\Drw$ and $\Delta\kappase$ oppose each other in Equation~\ref{eq:sin}
reduces the sensitivity of $\swsqeffl$ to the corrections.
The effect of $\Drw$ is larger, so that changes in $\swsqeffl$ are of
the same sign as those in $\swsq$, but are a factor of $\swsq$/$\cwsq$
smaller in magnitude~\cite{Jegerlehner:1991ed}, which is about $0.3$.

Since $\swsq$ is present in the dominant, $\Mt$, term of
Equation~\ref{eq:deltakappa}, but absent in the corresponding terms of
Equations~\ref{eq:deltarho} and~\ref{eq:deltarhomhsmall}, the effects of
$\dalhad$ are felt in $\Delta\kappase$, but $\Delta\rhose$ is well isolated.
The uncertainty in $\dalhad$ therefore dilutes the inherent precision of the
$\swsqeffl$ measurements, but not those of the partial widths.

Similarly, the weak effects of $\Drw$ are felt undiluted in $\MW$, so
that the relative effect of $\Delta\alpha$ is much smaller there than it is
for $\swsqeffl$, where about $0.7$ of the weak corrections are canceled
through $\Delta\kappase$.

The flavour dependence of the radiative corrections is very small for all
fermions, except for the b quark, where vertex corrections are significant
due to the large mass
splitting between the bottom and top quarks~\cite{Jegerlehner:1991dq},
resulting in:
\begin{eqnarray}
  \label{eq:b_vertex}
  \Delta\rho_{\mathrm{b}}   &=& -\frac{\GF\Mt^2}{2\sqrt{2}\pi^2} + \cdots.
\end{eqnarray}
Notice that $\Delta\rho_{\mathrm{b}}$ has no Higgs dependence, making the
quantity $\Rbz = \Gbb/\Ghad$ a straight-forward indicator of $\Mt$, since the
$\MH$ dependences entering through $\Delta\rhose$ cancel in numerator and
denominator.

The partial Z decay widths contain final state radiative corrections and
add up straightforwardly to yield the total width of the $\Zzero$
boson~\cite{PCP99}:
\begin{eqnarray}
 \Gff  = && N_c^{\rm f} \frac {\GF \MZ^3} {6\sqrt{2}\pi} 
      \left( |\cgaf|^2 R_{\mathrm{Af}} + |\cgvf|^2 R_{\mathrm{Vf}} \right) 
      \nonumber \\ 
     &&+  \Delta_{\rm ew/QCD}. 
\label{eq:Gff}
\end{eqnarray} 
Here the complex effective couplings, $\cgvf$ and $\cgaf$ differ only
by small imaginary parts from the real effective couplings defined
in Equations~\ref{eq:gveff} and~\ref{eq:gaeff}.
The radiator factors $R_{\mathrm{Vf}}$ and $R_{\mathrm{Af}}$ take into
account final state QED and QCD~\cite{bib-PCLI-QCD} corrections
including pair production; $\Delta_{\rm ew/QCD}$ accounts for small
contributions from non-factorizable electroweak/QCD corrections.

Although the discussion of electroweak corrections presented above
(Equations~\ref{eq:deltakappa}--\ref{eq:b_vertex}) is at the
single loop level, serious quantitative calculations implemented
in programs such as ZFITTER and TOPAZ0 include terms of
higher order.
These programs can not only calculate the effective couplings as a
function of the free Standard Model parameters, but can also relate
these couplings to observables such as cross sections and partial
widths.
The simple single-loop expressions are nevertheless useful in
understanding the results of the elaborate, very precise calculations
used in deriving our quantitative results.

\section{$\Zpole$ Constraints on $\Mt$ and $\MH$}
                 
As $\Zpole$ input data we consider the set listed in Table 16.1 of
reference~\cite{Group:2003ih}, which consists of $\dalhad$, plus the 14
$\Zpole$ measurements: $\MZ$, $\GZ$, $\shad$, $\Rl$, $\Afbzl$,
$\cAl~(P_\tau)$, $\swsqeffl$($\avQfb$), $\cAl$~(SLD), $\Rbz{}$, $\Rcz{}$,
$\Afbzb{}$, $\Afbzc{}$, $\cAb$, and $\cAc$, including their error
correlations.  We perform Standard Model fits\footnote{Our results are
equivalent to those described in~\cite{Group:2003ih}. The resulting
numerical values for the Standard Model parameters can be found in
Table 16.2 of that reference.} to this data, using ZFITTER~6.36 to predict the
observables as a function of the five free parameters: $\dalhad$, $\MZ$,
$\alfas$, $\Mt$ and $\MH$.

To illustrate the constraints imposed by these measurements, it is useful to
group them according to their functional roles.
Two quantities, $\dalhad$ and $\MZ$, appear as both input data and
fit parameters.
The electromagnetic coupling, parametrized through $\dalhad$, is determined
externally~~\cite{bib-BP01}, and passes through the $\Zpole$ fit as an
inert ingredient.
Similarly, $\MZ$ is determined with such precision that it is incapable of
being pulled by its dynamic relation to other quantities.
Both might just as well have been treated as external constants, like
$\GF$.
The b- and c-quark asymmetry parameters, $\cAb$, and $\cAc$ are also inert in
the fit, since they have essentially no dependence on the Standard Model
parameters.

The three measurements $\GZ$, $\shad$ and $\Rl$ are equivalent to the more
meaningful, but more correlated quantities $\Gll$, $\Ghad$ and $\Ginv/\Gll$.
Of these, the latter is insensitive to variations in the Standard Model
Parameters, and serves to measure the number of light neutrino generations.
The strong coupling constant, $\alfas$, is determined by its effect on the
hadronic width, $\Ghad$, leaving $\Gll$ as the key quantity in constraining
$\Delta\rhose$.
The small size of $\gvl$ compared to $\gal$ (see Equation~\ref{eq:gveff})
makes $\Gll$ essentially independent of $\swsqeffl$.

As already discussed in connection with Equation~\ref{eq:b_vertex}, $\Rbz$
serves as a direct indicator of $\Mt$.
The role of $\Rcz$ is diffuse and non-critical.

The six remaining measurements, $\Afbzl$, $\cAl~(P_\tau)$,
$\swsqeffl$($\avQfb$), $\cAl$~(SLD), $\Afbzb{}$, and $\Afbzc{}$ serve to
determine the single quantity $\swsqeffl$.
Although there is a disturbing lack of consistency between the values of 
$\swsqeffl$ derived from the two most precise measurements, $\cAl$~(SLD) and
$\Afbzb{}$, the discrepancy ($\approx 2.9 \sigma$)\footnote{the $\Afbzb{}$
measurements are not yet final} between them appears to elude explanation.
Within the Standard Model all six of these measurements, and particularly the
two most precise, appear to be valid and well-defined measurements of
$\swsqeffl$.

Both measurements claim to be dominated by statistical uncertainties.
The measurements~\cite{Group:2003ih} of $\Afbzb{}$ are complex, but the level
of agreement between
the independent analyses of the four LEP Collaborations is excellent.
Common QCD corrections play a role, but are believed to be well understood.
The SLD measurement~\cite{ref:sld-al2000} of $\cAl$ through $\ALR$, the
asymmetry between the interaction rate for right- and left-handed electron
beam polarizations, is both simple and elegant.
The least implausible source of systematic error, in the measurement of the
beam polarization, is believed to be small and well-controlled.

Here we investigate the consequences of excluding either $\cAl$~(SLD) or
$\Afbzb{}$ from the fit, but we make no attempt to choose between them.
Either exclusion option restores acceptable consistency to the set of
$\swsqeffl$ measurements, but obviously introduces a bias.

The constraints on $\Mt$ and $\MH$ imposed by the 14 $\Zpole$ measurements
can therefore be almost
completely expressed in terms of three quantities: 
\vspace*{0.5cm}

$\Gll$ - related directly to $\Delta\rhose$

$\swsqeffl$ - related directly to $\Delta\kappase$, and

$\Rbz$ - related directly to $\Mt$.

\subsection{$\Zpole$ Data Constraints Alone}

\begin{figure}[ht]
  \begin{center}
  \ifthenelse{\equal{color}{\color}}{
    \includegraphics[width=1.0\columnwidth]{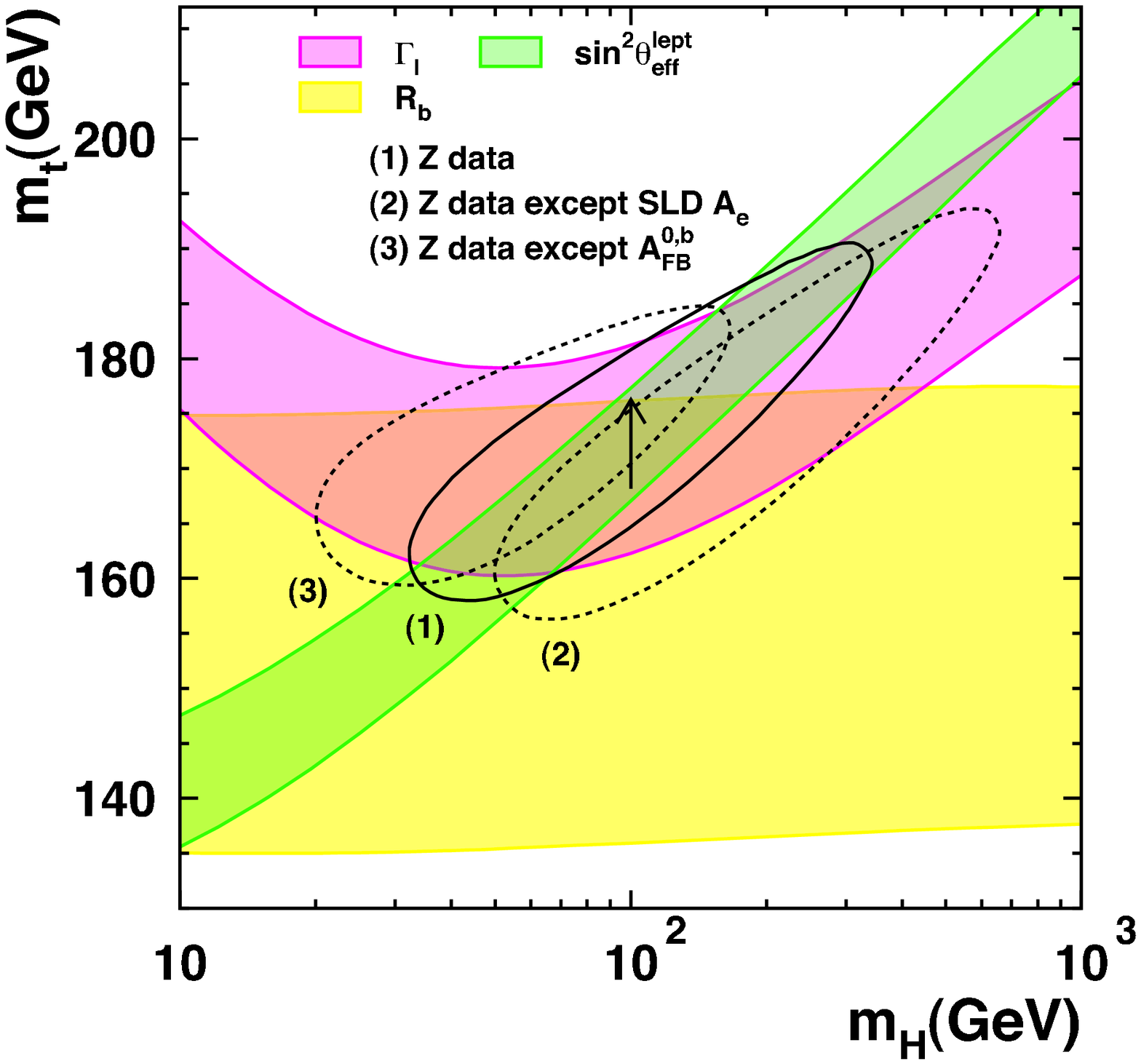}}{
    \includegraphics[width=1.0\columnwidth]{mtvsmh-hs0}
    }
    \caption[$\Mt$ and $\MH$ constraints from the Z-pole measurements]
    { $\Mt$ and $\MH$ constraints from the Z-pole measurements. Each band gives
    the $\pm1\sigma$ constraint from the indicated measurement.  The arrow indicates the
    additional variation in $\swsqeffl$ due to a $\pm1\sigma$  uncertainty
    in $\dalhad$.  The ellipses give the 68\% CL fit contours for the indicated
    data.}
    \label{Mt_vs_Mh}
  \end{center}
\end{figure}

How these components of the Z-pole data
constrain $\Mt$ and $\MH$ can be clearly illustrated on a plot of $\Mt\,vs\,
\LOGMH$, as shown in Figure~\ref{Mt_vs_Mh}.
Shown here are the usual 68\% CL error contours, plotted by finding the curve
in the $\Mt\,vs\,\LOGMH$ plane where the $\chi^2$ of the fit exceeds its minimum
by 2.28.
For projections in a single parameter, 68\% CL corresponds to $1.51\sigma$.
Also plotted are measurement constraint bands, which show the contribution of
each measurement.
These are plotted by finding the curves in the $\Mt\,vs\,\LOGMH$ plane where the
predicted value of the measured quantity equals the central value of the
measurement $\pm1\sigma$.
Note that since each of these bands indicate the constraint imposed by a fixed
measurement, it exhibits exactly the inverse parameter dependence compared to
the Standard Model prediction for the measured quantity.

\vspace{0.3cm}
The diagonal band in Figure~\ref{Mt_vs_Mh} indicates the $\pm1\sigma$ constraint from the
$\swsqeffl$ measurements.  Higher-order electroweak corrections do not
significantly affect the linear relationship expected from the lowest-order
terms, and the slope of this band is approximated by the ratio of the
coefficients of $\Mt^2$ and $ln(\MH)$ in the expression for $\Delta\kappase$\footnote{
Note that the success of this approximation relies on $\Drw$ remaining constant
along lines of constant $\swsqeffl$, as mentioned in the  
text discussing Equation~\ref{eq:kappadrw}.}:
\begin{eqnarray}
   \label{eq:kappaslope}
    \left.  \frac{d\Mt}{d ln(\MH)}\right|_{\kappase}  &=&  \frac{10}{9}\frac{
\MZ^2 \swsq}{ \Mt}  .
\end{eqnarray}

\vspace{0.3cm}
The banana-shaped band in Figure~\ref{Mt_vs_Mh} shows the $\pm1\sigma$ constraint from the $\Gll$
measurement.  The slope of this band at large $\LOGMH$ agrees reasonably with the
linear relation expected from the ratio of coefficients in the lowest-order
expression for $\Delta\rhose$ in this region:
\begin{eqnarray}
   \label{eq:rhoslope}
    \left.  \frac{d\Mt}{d ln(\MH)}\right|_{\rhose}  &=& \frac{ \MZ^2
}{\Mt}\swsq . 
   \end{eqnarray}
\noindent
Also at low $\LOGMH$ the slope, now negative, agrees  with the
lowest-order expression given in Equation~\ref{eq:deltarhomhsmall}:
\begin{eqnarray}
   \label{eq:rhomhsmalls} 
    \left.  \frac{d\Mt}{d ln(\MH)}\right|_{\rhose}  &=& - \frac{2 \MZ^2}{3\Mt} .
\end{eqnarray}

\vspace{0.3cm}
Due to the fact that $\Rbz$ is controlled by vertex corrections determined by
the $t-b$ mass-splitting, its measurement provides a constraint on $\Mt$
almost independent of $\LOGMH$, as shown by the broad horizontal band in
Figure~\ref{Mt_vs_Mh} (also $\pm1\sigma$).

\vspace{0.3cm}
The central ellipse in Figure~\ref{Mt_vs_Mh} shows the 68\% CL contour for
the Standard Model fit to
all Z-pole measurements.  The dominant role played by $\swsqeffl$ in
determining the minor axis is evident, and the turn-over of the $\Gll$ banana
provides the lower bound of the major axis.  The $\Rbz$ measurement provides
the upper bound.  If the $\Rbz$ constraint is removed, the fit in fact no
longer yields a 68\% CL upper limit for $\MH$ or $\Mt$.

\vspace{0.3cm}
The other two ellipses in Figure~\ref{Mt_vs_Mh} show similar fit contours
when either the $\ALR$ or the
$\Afbzb$ measurements are excluded from the fit.  The noticeable shrinkage of
the major axis in the case when the $\Afbzb$ measurement is dropped is due to
the fact that the $\swsqeffl$ constraint then begins to move around the corner
of the $\Gll$ banana.  If the $\swsqeffl$ measurement moved even lower in $\MH$, the
$\swsqeffl$ and $\Gll$ constraints would become almost perpendicular,
eliminating the usual $\Mt - \LOGMH$ error correlation.

\vspace{0.3cm}
The agreement of  the indirect measurement of $\Mt$ from the Z-pole
measurements alone with the direct measurement made in $\pp$
collisions~\cite{PDG2002} ($\Mt = 174.3 \pm 5.1$ GeV) is an important
experimental confirmation of the validity of electroweak corrections.  The
remarkable stability of the indirect measurement's central value under shifts
in $\swsqeffl$ can be seen to result from a complex interplay between the
relatively weak constraint from $\Rbz$, which happens to lie low, and the exact
relation between the $\swsqeffl$ measurement band and the position of the
$\Gll$ corner.

\vspace{0.3cm}
The arrow in Figure~\ref{Mt_vs_Mh} shows how the $\swsqeffl$ measurement
band would shift under a
$\pm 1 \sigma$ change in the $\dalhad$ determination.  Only the $\swsqeffl$
measurement is significantly sensitive to $\dalhad$ on the scale of the
current  errors, making the effective width of the $\swsqeffl$ measurement
band in the $\Mt\, vs\, \MH$ plane about 50\% wider than the band shown, which
corresponds to  $\dalhad$ fixed at its central value.

\vspace{0.3cm}
The effect of applying the constraint of the direct $\Mt$ measurement~\cite{PDG2002} can
easily be visualized by imagining a horizontal band at $\Mt =  174.3 \pm 5.1$ GeV. 
Notice that at the operating point of the Z-pole fit, the direct $\Mt$
measurement essentially surplants not only $\Rbz$, but also the constraints
provided by $\Gll$.

\vspace{0.3cm}
It is perhaps interesting to remark on the fact that all measurements are
compatible with the broad $\Gll$ extremum in $\LOGMH$.  Only the failure to
find direct production of the Higgs boson at LEP II indicates nature's choice to lie
on the $\Gll$ upper branch.

\subsection{Additional Constraints from $\MW$}

Direct measurements of $\MW$ also provide important constraints on the Standard Model
parameters through the electroweak corrections $\Drw$.
The high precision of the $\MZ$ measurement means that the constraint imposed on $\swsq$ through
Equation~\ref{eq:rho} is effectively limited only by the measurement uncertainty in $\MW$.
Although further precision on $\MW$, from both LEP II~\cite{Group:2003ih} and the
Tevatron~\cite{PP-MW-GW:combination} can be expected, the available combined result,
$\MW=80.426\pm0.034~\GeV$~\cite{Group:2003ih}
is already sufficient to provide
constraints on $\Mt$ and $\MH$ which are comparable in precision to those derived from
the $\Zpole$ alone.
The approximate equivalence $\Drw \approx -\Delta\kappase$ of
Equation~\ref{eq:kappadrw} becomes less exact towards small values of $\MH$, so that
the direct $\MW$ measurements allow some separation of Higgs and top effects when combined
with the $\Zpole$ measurement of $\swsqeffl$.

\begin{figure}[ht]
  \begin{center}
  \ifthenelse{\equal{color}{\color}}{
    \includegraphics[width=1.0\columnwidth]{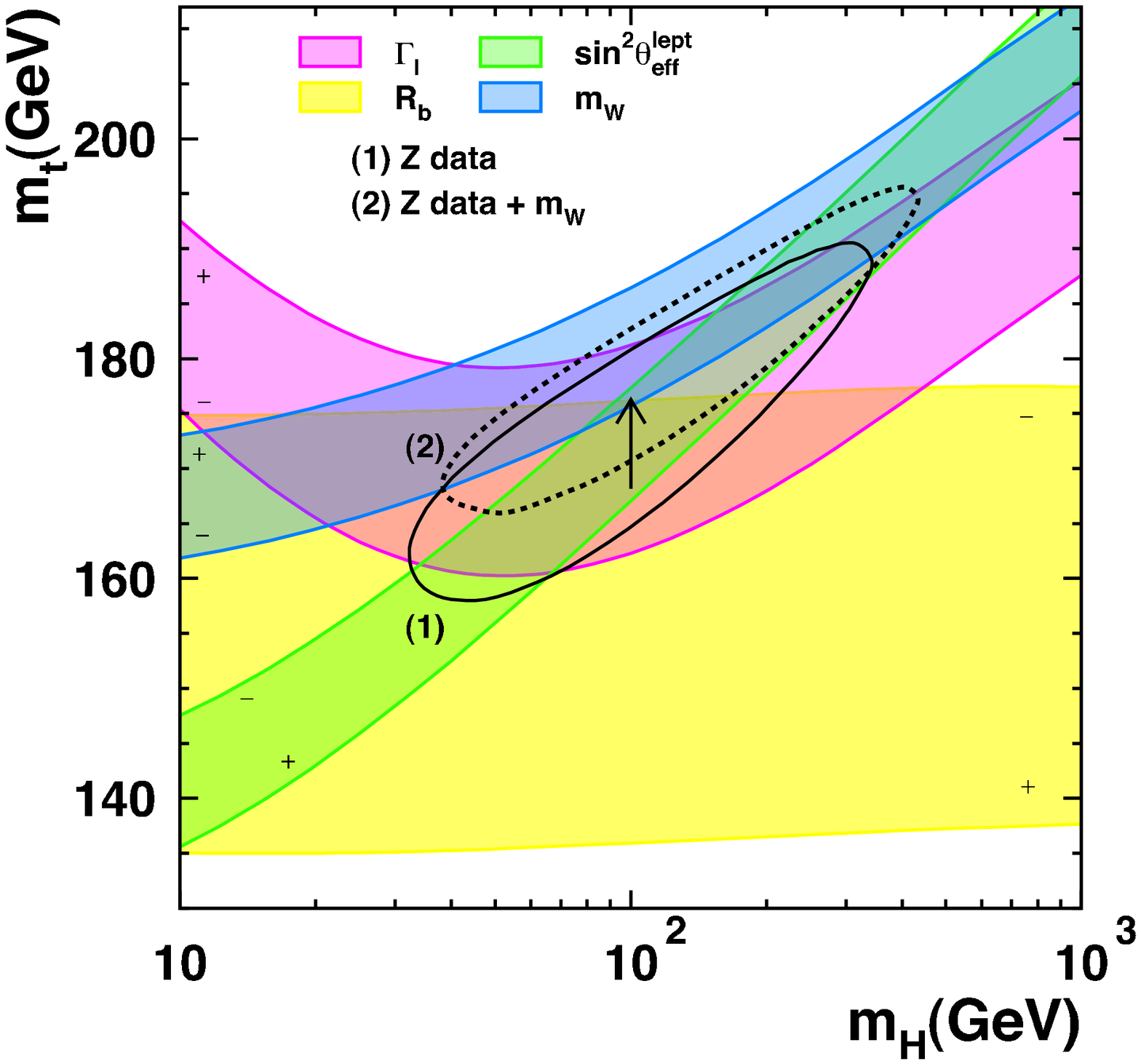}}{
    \includegraphics[width=1.0\columnwidth]{mtvsmh-hs0mwcont}
    }
    \caption[$\Mt$ and $\MH$ constraints from Z-pole measurements and $\MW$]
    { $\Mt$ and $\MH$ constraints from Z-pole measurements and $\MW$.
    Each band gives the $\pm1\sigma$ constraint from the indicated measurement.
    The $\pm$ signs on each band show the sense of variation in the measurement.
    The arrow indicates the
    additional variation in $\swsqeffl$ due to a $\pm1\sigma$  uncertainty
    in $\dalhad$.
    The error contours are shown at 68\% CL for fits to both the $\Zpole$ results alone (solid),
    and the $\Zpole$ results combined with $\MW$ (dashed).}
    \label{Mt_vs_Mhmw}
  \end{center}
\end{figure}

\vspace{0.3cm}
The diagonal $\MW$ band in Figure~\ref{Mt_vs_Mhmw} shows the constraint from the
preliminary direct measurements of $\MW$.
The flattening of its slope at low values of $\MH$, and the fact
that it happens to lie relatively higher in $\Mt$ than the $\swsqeffl$ band,
leads to a clear divergence of the two measurement bands at low $\MH$.

\vspace{0.3cm}
Compared to the solid error contour of the $\Zpole$ fit
without $\MW$, which is shown again in Figure~\ref{Mt_vs_Mhmw}, the dashed error contour
which includes the
$\MW$ measurements is displaced along the converging $\MW$ and $\swsqeffl$ measurement
bands like the slider of a zipper towards larger values of $\Mt$ and $\MH$.  This
demonstrates that the $\MW$ measurement provides a stronger lower bound along this
axis than does $\Gll$.

\vspace{0.3cm}
The contour including $\MW$ is also displaced perpendicularly and is therefore more
compatible with the lower value of $\swsqeffl$ derived from $\ALR$ than the higher value
favored by $\Afbzb$ (see Figure~\ref{Mt_vs_Mh}).  

\vspace{0.3cm}
Without constraining $\Mt$ it is the first displacement, parallel to the $\swsqeffl$
band which dominates, and the addition of $\MW$ shifts the favored range for $\MH$
upwards, as already mentioned, with respect to the $\Zpole$-only fit.

\vspace{0.3cm}
When the direct measurement of $\Mt = 174.3 \pm 5.1$ GeV is imposed at the
outset, however, not only is $\MH$ constrained more tightly, but the second,
perpendicular displacement of the error ellipse then dominates, and the
addition of $\MW$ shifts the favored range for $\MH$ downwards, rather than
upwards.


\section{Higgsstrahlung}

\label{sec:higgsstahlung}
\begin{figure}[ht]
  \begin{center}
    \includegraphics[width=0.4\columnwidth]{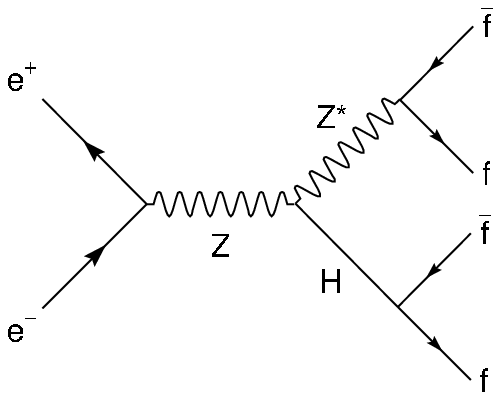}
    \caption[Higgsstrahlung at the $\Zpole$]
    {Higgsstrahlung at the $\Zpole$. The narrow width of the Higgs boson
    implies that its decay products have an invariant mass near $\MH$, while the
    $\Zzero^*$ is sufficiently off-shell to ensure energy conservation.}
    \label{feynmann}
  \end{center}
\end{figure}

Since direct searches for the Higgs boson have demonstrated that $\MH > 114~\GeV$\,
at 95\% CL~\cite{LEPSMHIGGS}, Figures~\ref{Mt_vs_Mh} and~\ref{Mt_vs_Mhmw}
and Equation~\ref{eq:deltarhomhsmall} follow
the precision calculations of ZFITTER and TOPAZ0 in neglecting the
Higgsstrahlung process shown in Figure~\ref{feynmann}.
However, the decrease of $\Delta\rhose$ at low $\MH$ predicted by
Equation~\ref{eq:deltarhomhsmall} is due to the effect of virtual Higgs loops.
These should properly be compensated by external Higgs corrections, which will
enter with opposite sign.
When choosing to ignore the direct Higgs search results to explore what the
$\Zpole$ measurements alone tell us about the Standard Model, it is logically
inconsistent to consider virtual Higgs corrections while neglecting the
concomitant external corrections which the Standard Model predicts for low
values of $\MH$.

\begin{figure}[ht]
  \begin{center}
    \includegraphics[width=1.0\columnwidth]{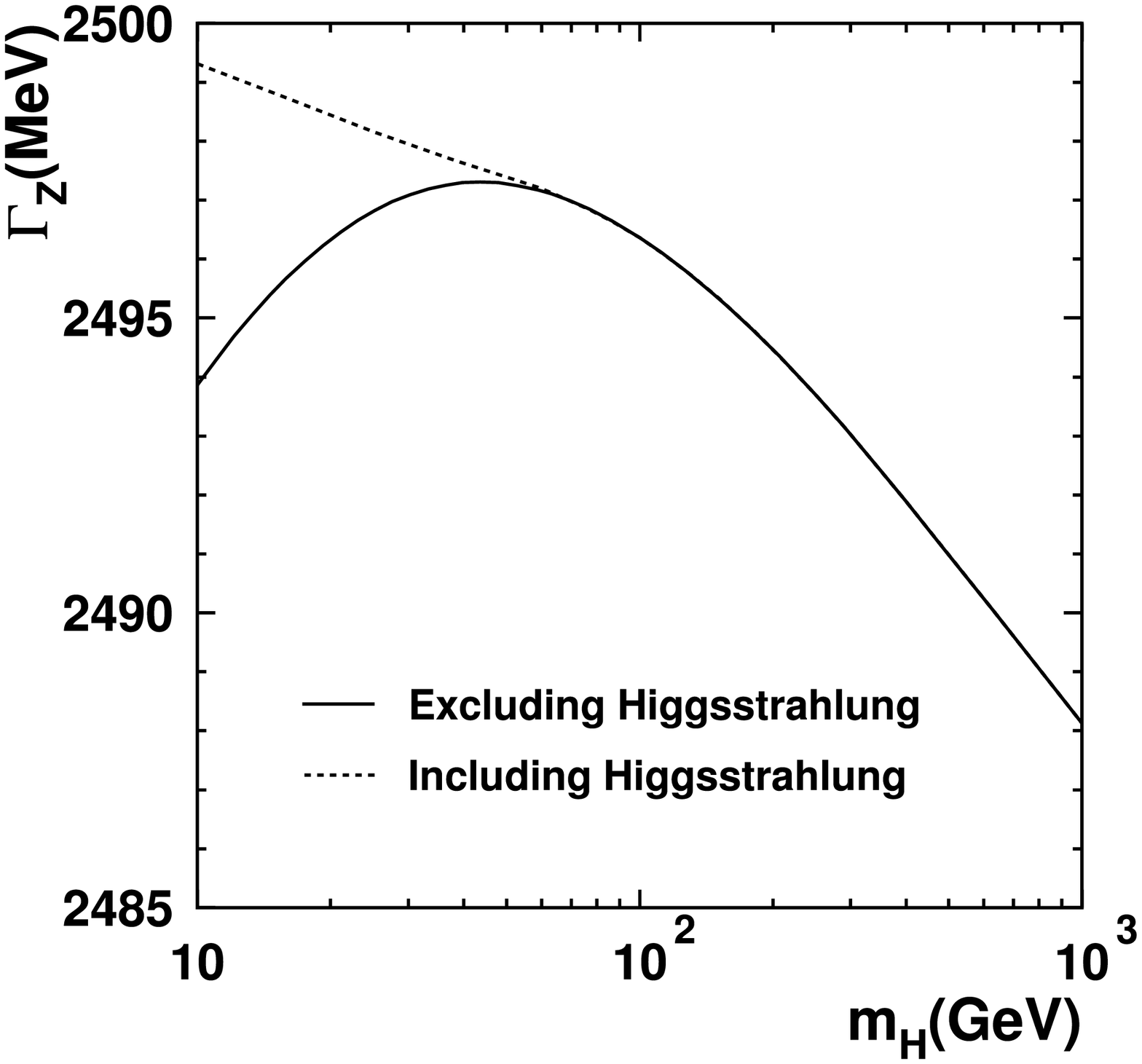}
    \caption{$\GZ\,vs\,\LOGMH$, including and excluding external
Higgsstrahlung.}
    \label{gamma_tot}
  \end{center}
\end{figure}

We have therefore undertaken calculations which account for the effects of the
expected Higgsstrahlung at low $\MH$. 
The fractional rate of such Higgsstrahlung is given in~\cite{LEP1YR89VOL2}:
\begin{eqnarray}
&&\frac{\Gamma(\Zzero \rightarrow H \,\ff)}{\Gamma(\Zzero \rightarrow \ff)} =
\frac{\alpha}{4\,\pi \, \cwsq\,\swsq} \times \nonumber\\
&&\int_{2r}^{1+r^2}\frac{\left( 1 +
\frac{2\,r^2}{3} - x + 
      \frac{x^2}{12} \right) \,{\sqrt{ x^2-4\,r^2}}}
  { \left( \frac{{\GZ}^2}{{\MZ}^2} + 
      {\left( x -r^2 \right) }^2 \right) } dx \,,
\label{eq:higgsstrahlung}
\end{eqnarray}
where
\begin{eqnarray}
x &=& 2 E_H / \MZ \,,\\
r &=& \MH / \MZ \,,\\
E_H &=& \frac{\MZ^2 + \MH^2 - m^2_\ff}{2\MZ} \,,
\end{eqnarray}
and $m_\ff$ is the invariant mass of the $\Zzero^*$ decay products.
The Higgsstrahlung of Equation~\ref{eq:higgsstrahlung} represents extra width
for $\Zzero$ decay, which is not included in ZFITTER.
Figure~\ref{gamma_tot} shows the total width of the $\Zzero$ with and without
this external Higgsstrahlung.
At values of $\MH$ below $\sim 50~\GeV$, which are still consistent with the
$\Zpole$ measurements, the Standard Model Higgsstrahlung would have been
appreciable, representing on the order of an $\MeV$ in the $\Zzero$ decay
width.

In order to incorporate the effects of such Higgsstrahlung into the $\Zpole$
analysis, we re-interpret the experimental measurements, which we consider
to include Higgsstahlung, in terms of the theoretical quantities as calculated
by ZFITTER, without Higgsstrahlung.
In the case of the total width this is particularly straight-forward:
\begin{equation}
\label{eq:gtoths}
\GZ^{\mathrm{meas}} = \GZ^{\mathrm{zf}} + \GZ^{\mathrm{hs}}
\end{equation}
where the superscripts have the obvious meanings: zf = ZFITTER, meas =
measured, hs = Higgsstrahlung.  Here $\GZ^{\mathrm{hs}} = \Gamma(\Zzero
\rightarrow H \,\ff)$, summed over all fermion species, is calculated
according to Equation~\ref{eq:higgsstrahlung}.
Since $\GZ$ is the resonance width of the $\Zzero$, determined by the decay
lifetime and the uncertainty principle, its measurement is entirely
independent of any details of the experimental event selection and analysis.

The manner in which the measurement of the partial widths will be affected is
less obvious.
Although experimental details differ among the four LEP collaborations, we
take OPAL as a representative experiment~\cite{Ahmet:1991eg,Abbiendi:2000hu}
and study how Higgsstrahlung would have affected the measurements.
Although it would be better to pursue these studies using detailed simulations
of all four experiments, the precision  which is necessary is sharply reduced
by a very
simple consideration: any Higgsstrahlung events classed as hadronic $\Zzero$
decays will simply increase the apparent value of $\alfas$, and will otherwise
not affect the analysis.
This is the case since, while $\Ghad$ and $\Gll$ share the same propagator
corrections through $\Delta\rhose$, final-state QCD radiation increases
the hadronic width alone by the factor $R_{\mathrm{QCD}}$:
\begin{equation}
  \label{eq:QCDcorr}
                    R_{\mathrm{QCD}} = 1 + \frac{\alfas}{\pi} + \cdots .
\end{equation}
Any shift in the $\Ghad/\Gll$ ratio is therefore interpreted by the fit as a
change in $\alfas$.

All four LEP experiments followed the natural strategy of essentially, if not
formally, accepting
all $\Zzero$ decays as hadronic unless they passed the stringent requirements
to be identified as one of the three species of leptons.
To an excellent approximation all Higgsstrahlung events in which the
$\Zzero^*$ decays hadronically will remain classed as hadronic events.
Since
the dominant branching fraction
of the Higgs boson in the relevant mass range
is to $\bb$, many other Higgsstrahlung events, except for very low values of
$\MH$, will also appear to be hadronic decays.


To clarify the experimental classification of the Higgsstrahlung events in
which the $\Zzero^*$ decays leptonically we studied samples of fully simulated
Higgstrahlung events in the OPAL detector.
These samples were generated using the program HZHA-03~\cite{hzha}, included
all accessible Higgs decays according to the Standard Model, and covered the
$\Zzero^*$ decay channels $\ee$, $\mumu$ and ${\antibar{\nu}}$.
Since $\Zzero^*$ decays to $\tautau$ were not available, we treated these as
an average of $\ee$ and $\mumu$.

For each $\Zzero^*$ decay channel, as a function of $\MH$, we calculated the
probability that the Higgstrahlung events would be classified as hadronic
$\Zzero$ decays.
If an event failed the hadronic selection, we assumed that it would be
classified correctly according to the actual $\Zzero^*$ decay.
Since the Standard Model assumes lepton universality, any cross-over
between charged lepton channels would in any case be irrelevant. 

We then calculated, as a function of $\MH$, two fractions: 

\vspace*{0.5cm}
\noindent
The fraction of Higgsstrahlung events, $\mathrm{F}^{had}_{\ell\ell}(\MH)$, in
which the $\Zzero^*$ in fact decayed to charged leptons, but the event was
identified as an hadronic $\Zzero$ decay.

\vspace*{0.5cm}
\noindent
The fraction of Higgsstrahlung events,
$\mathrm{F}^{had}_{\mathrm{inv}}(\MH)$,in which the $\Zzero^*$ in fact decayed
to neutrinos, but the event was identified as an hadronic $\Zzero$ decay.

\begin{figure}[ht]
  \begin{center}
    \includegraphics[width=0.49\columnwidth]{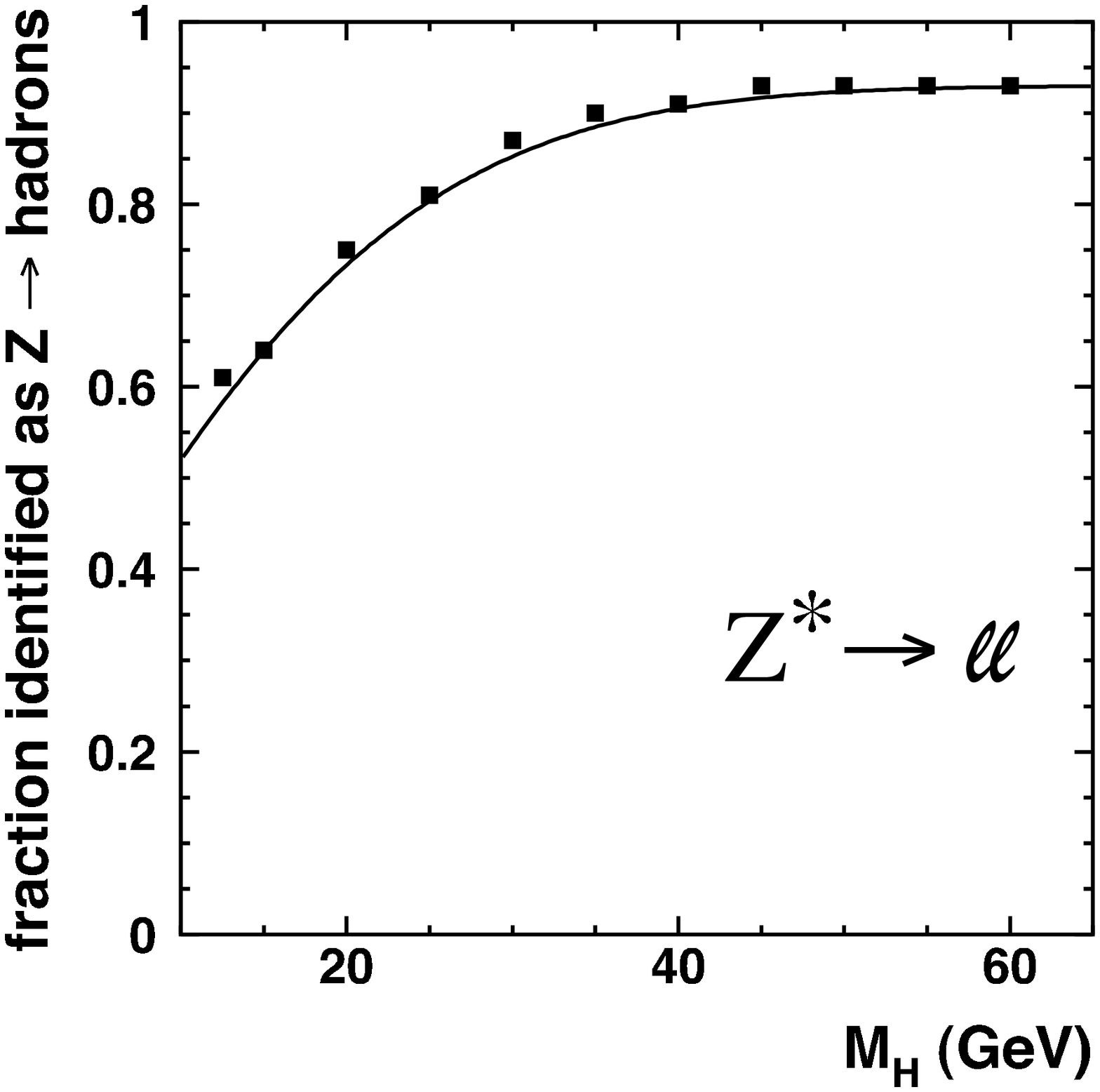}
    \includegraphics[width=0.49\columnwidth]{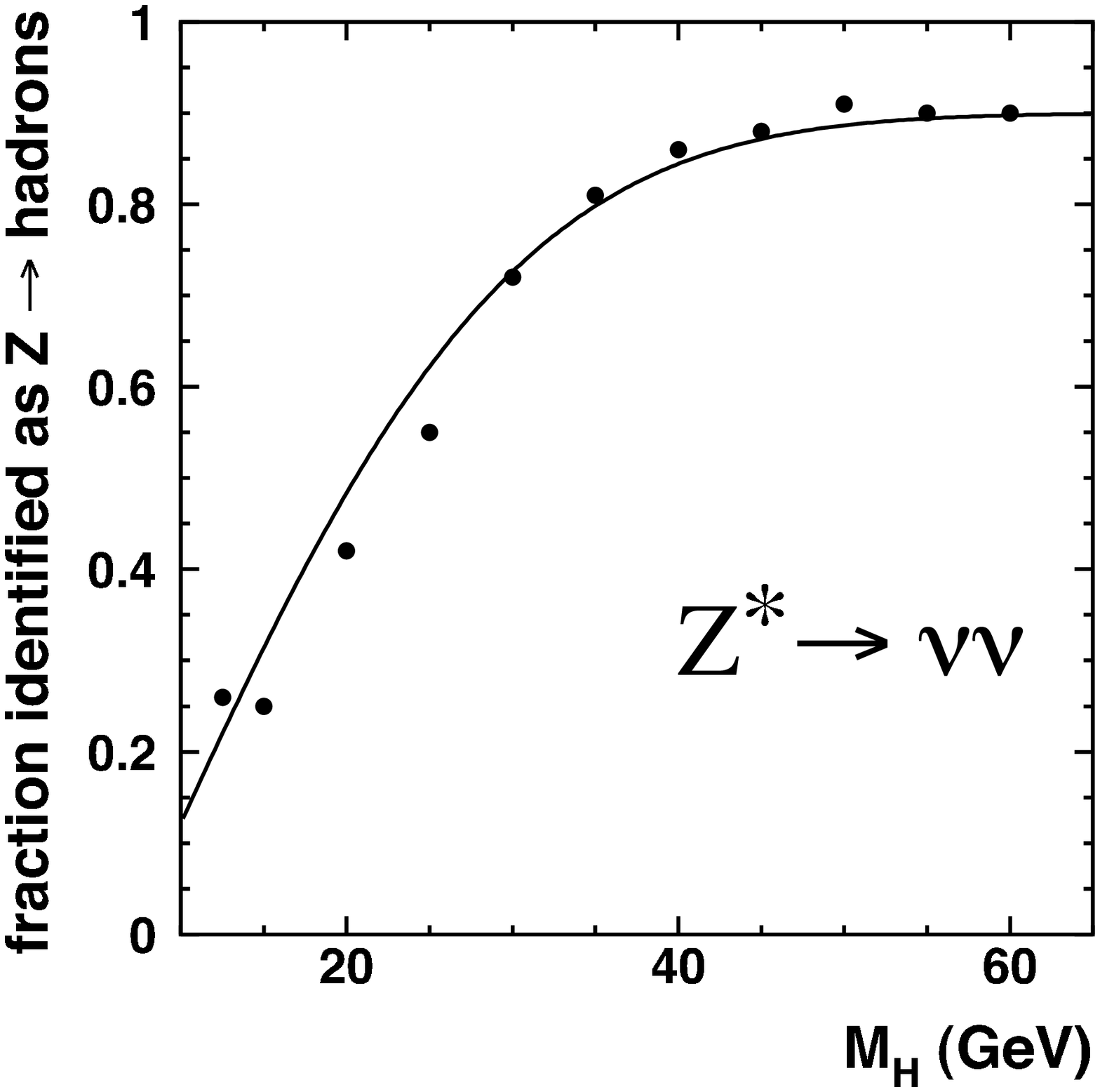}
    \caption[The fractions of Higgstrahlung events clasified as hadronic
    $\Zzero$ decays]
    {As a function of $\MH$, the fractions of Higgsstrahlung events
    where the $\Zzero^*$ in fact decayed, respectively, to charged leptons or
    neutrinos, but the event was identified as an hadronic $\Zzero$ decay.}
    \label{fig:zll}
  \end{center}
\end{figure}

\vspace*{0.5cm}
\noindent
Figure~\ref{fig:zll} displays these fractions along with a smooth function
fitted to represent the Monte Carlo data.

We then calculate the measured  partial widths, including the effects of
Higgsstrahlung, using the smooth $\mathrm{F}^{had}_{\mathrm{xx}}$ functions:

\begin{eqnarray}
\label{eq:gllhs}
\Ghad^{\mathrm{meas}} &=& \Ghad^{\mathrm{zf}}
+ \mathrm{F}^{had}_{\ell\ell}(\MH) \Gll^{\mathrm{hs}}
+ \mathrm{F}^{had}_{\mathrm{inv}}(\MH) \Ginv^{\mathrm{hs}} \,,\\
\label{eq:ghadhs}
\Gll^{\mathrm{meas}} &=& \Gll^{\mathrm{zf}}
+ \left[ 1-\mathrm{F}^{had}_{\ell\ell}(\MH) \right] \Gll^{\mathrm{hs}} \,,\\
\label{eq:ginvhs}
\Ginv^{\mathrm{meas}} &=& \Ginv^{\mathrm{zf}}
+ \left[ 1 - \mathrm{F}^{had}_{\mathrm{inv}}(\MH) \right] \Ginv^{\mathrm{hs}}\,.
\end{eqnarray}

We take the effect of Higgsstrahlung on the measurement of $\swsqeffl$ to be
negligible.  Clearly there is no effect on the SLD measurement of $\ALR$, since
it concerns only the initial state.
Similarly, there can be no effect on the average $\tau$ polarization. The
asymmetry measurements have a relative precision of at best a few percent, and
any possible dilution of the asymmetry due to kinematic distortions in the few
$10^{-3}$ fraction of Higgstrahlung events will not be significant.

\begin{figure}[ht]
  \begin{center}
    \ifthenelse{\equal{color}{\color}}{
    \includegraphics[width=1.0\columnwidth]{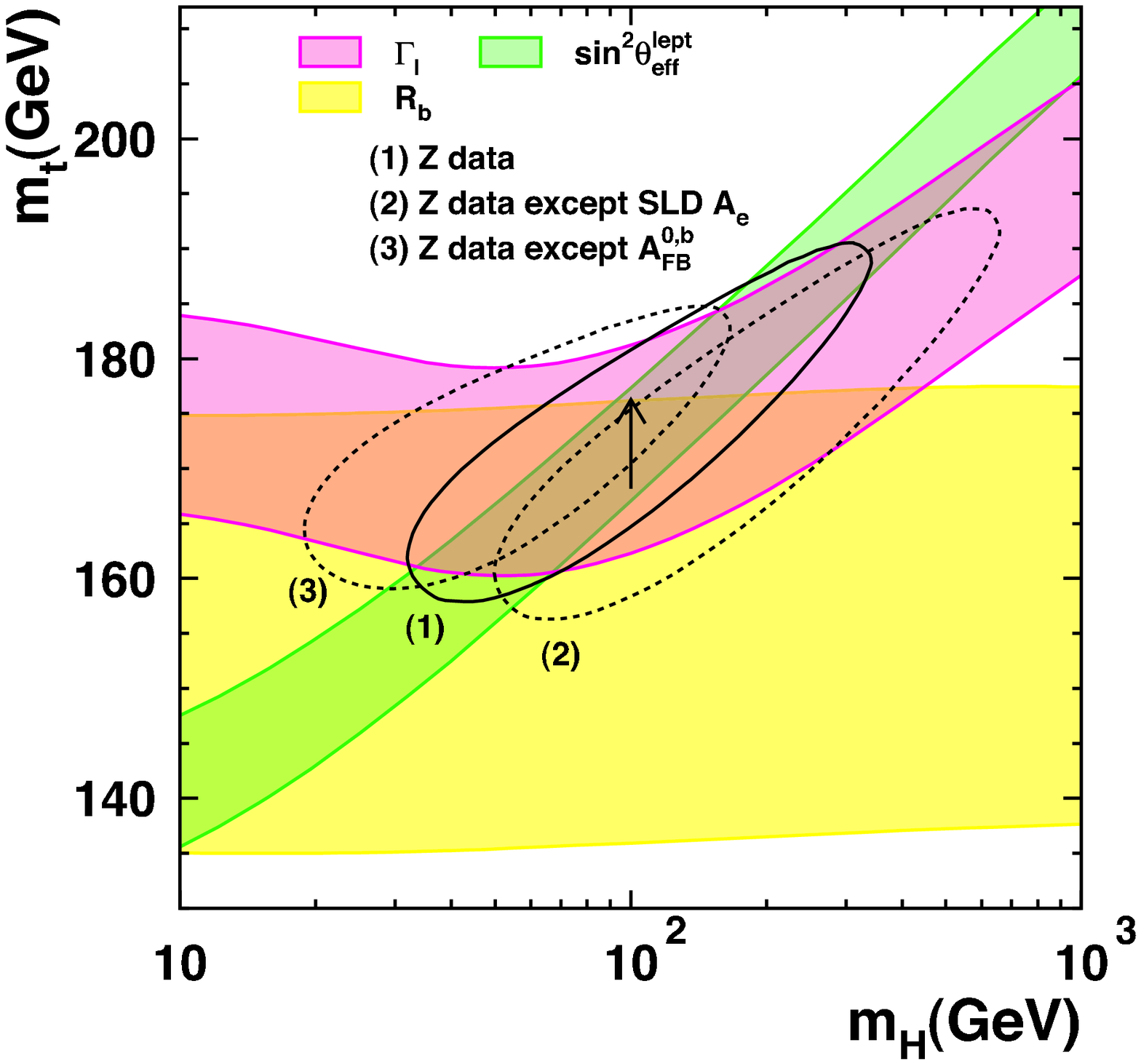}}{
    \includegraphics[width=1.0\columnwidth]{mtvsmh-hs5}
    }
    \caption[$\Mt$ and $\MH$ constraints from Z-pole measurements with
    Higgsstrahlung taken into account]
    {$\Mt$ and $\MH$ constraints from Z-pole measurements with
    Higgsstrahlung taken into account.  Each band gives
    the $\pm1\sigma$ constraint from the indicated measurement.  The arrow indicates the
    additional variation in $\swsqeffl$ due to a $\pm1\sigma$  uncertainty
    in $\dalhad$.  The ellipses give the 68\% CL fit contours for the indicated
    data.}
    \label{Mt_vs_Mh_hs5}
  \end{center}
\end{figure}

With the relations of Equations~\ref{eq:gtoths}
and~\ref{eq:gllhs}-\ref{eq:ginvhs}, we then re-made the plot of
Figure~\ref{Mt_vs_Mh}, using ZFITTER to calculate the measurement error bands
and 68\% CL contours for the re-intepreted measurements.
The result is shown in Figure~\ref{Mt_vs_Mh_hs5}.
Notice that the only visible change is the expected decrease in the slope of
the
$\Gll$ measurement band at low $\MH$, due to failure of some leptonic
Higgsstrahlung events to be identified as hadrons.  The lower extremity of the
68\% CL contour for the fit which drops the $\Afbzb$ measurement extends very
slightly, reflecting the local shift in the $\Gll$ constraint.
The lower $1\sigma$ error on $\Mt$ is only 2.4\% larger than it was when
Higgsstrahlung was ignored.
As a test of our proceedures, we also carried out the same calculations assuming
that all Higgsstrahlung was accepted as hadronic $\Zzero$ decay.
As expected, the resulting plot was indistinguishable from Figure~\ref{Mt_vs_Mh}.

\begin{figure}[ht]
  \begin{center}
    \ifthenelse{\equal{color}{\color}}{
    \includegraphics[width=1.0\columnwidth]{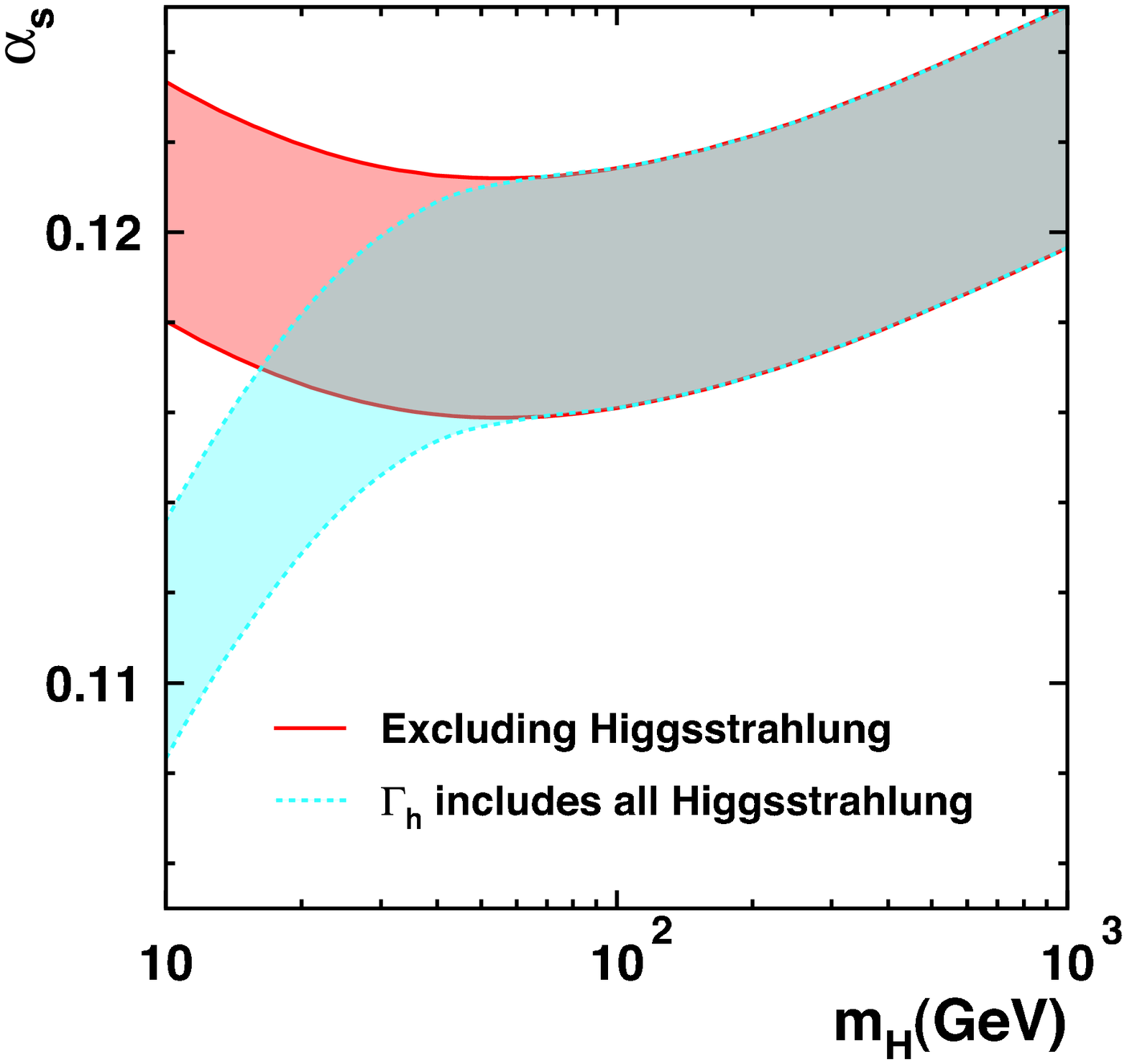}}{
    \includegraphics[width=1.0\columnwidth]{alfsvsmh}
    }
    \caption[The $\pm 1 \sigma$ error bands for $\alfas$]
    { The $\pm 1 \sigma$ error bands for $\alfas$ found in the four-parameter
    fits as a function of $\LOGMH$, with and without Higgsstrahlung.}
    \label{alfsvsmh}
  \end{center}
\end{figure}

In order to quantify the effect on $\alfas$ of Higgsstrahlung being identified
as hadronic $\Zzero$ decay, we performed a series of four-parameter fits, with
$\alfas$, $\alqed$, $\MZ$, and $\Mt$ free, for a series of fixed $\MH$ values.
We plot the $1\sigma$ error bands of $\alfas$ as a
function of $\LOGMH$ in Figure~\ref{alfsvsmh}, showing how
$\alfas$ shifts to compensate for the Higgsstrahlung which is
identified as hadronic $\Zzero$ decay.
Since $\Ghad^{\mathrm{zf}}$, the hadronic width as re-interpreted for ZFITTER,
is reduced from the fixed measured value, $\Ghad^{\mathrm{meas}}$, by the
Higgsstrahlung contribution, the hadronic final-state correction appears
smaller, and the fit value of $\alfas$ is reduced as $\MH$ becomes small.

\begin{figure}[ht]
  \begin{center}
    \ifthenelse{\equal{color}{\color}}{
    \includegraphics[width=1.0\columnwidth]{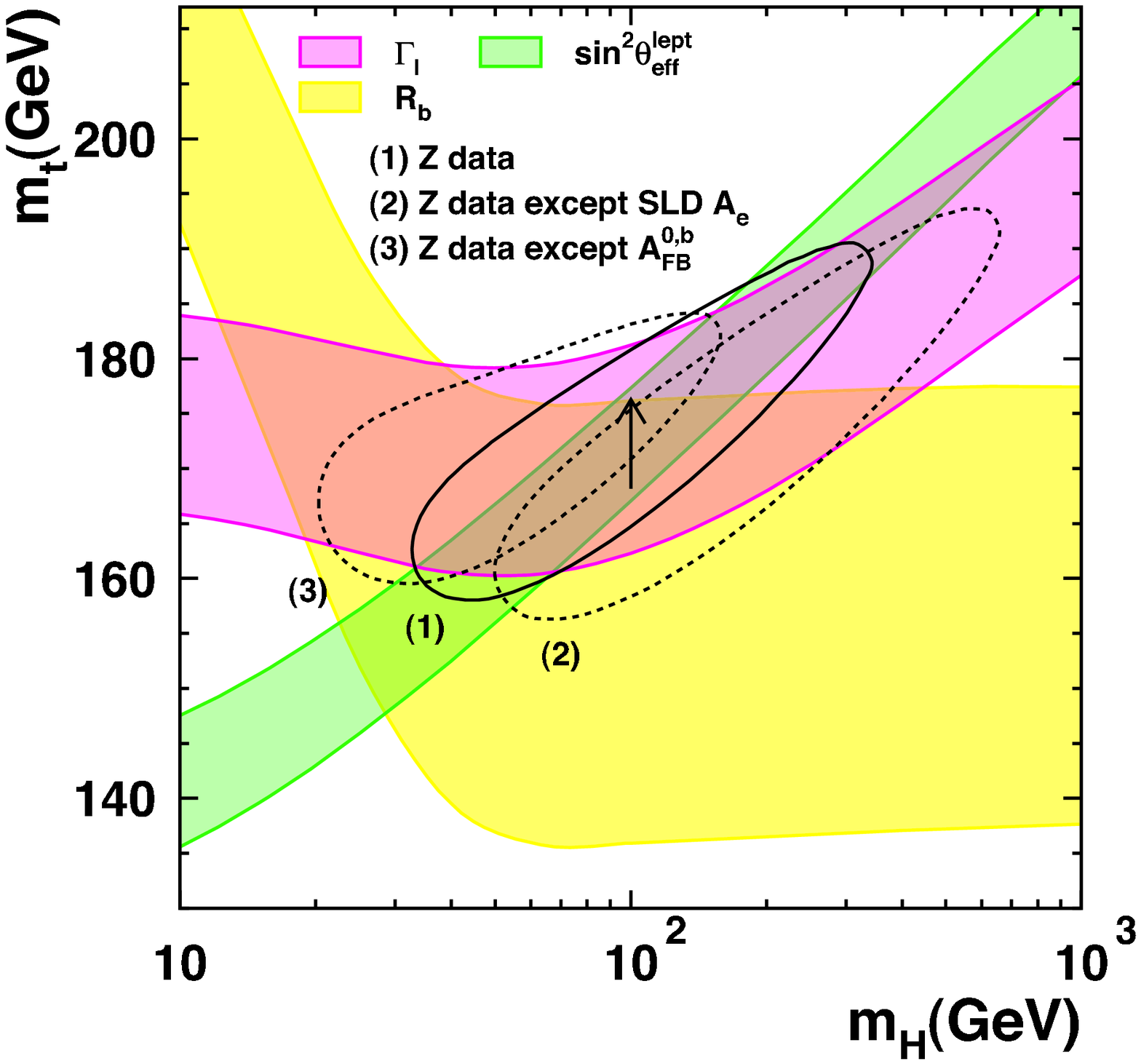}}{
    \includegraphics[width=1.0\columnwidth]{mtvsmh-hs6}
    }
    \caption[$\Mt$ and $\MH$ constraints assuming all
    Higgstrahlung identified as hadrons contributes to $\Gbb$]
    { $\Mt$ and $\MH$ constraints from Z-pole measurements with
    Higgsstrahlung taken into account.  Here the assumption is made that all
    Higgstrahlung identified as hadrons contributes to $\Gbb$.
    Each band gives
    the $\pm1\sigma$ constraint from the indicated measurement.  The arrow indicates the
    additional variation in $\swsqeffl$ due to a $\pm1\sigma$  uncertainty
    in $\dalhad$.  The ellipses give the 68\% CL fit contours for the indicated
    data.}
    \label{Mt_vs_Mh_hs6}
  \end{center}
\end{figure}

The $\Rbz$ measurement band remains unchanged in Figure~\ref{Mt_vs_Mh_hs5},
since we have assumed that the Higgstrahlung events identified as hadrons do not
disturb the quark flavor balance.
Actually, since the Higgs boson decays predominantly to $\bb$, it is likely that
Higgsstrahlung will increase $\Gbb$ proportionally more than the width of
other flavors.
A complete heavy flavor analysis was beyond the scope of this study, so we
considered the limiting case that all Higgsstrahlung events where the Higgs boson
decays to $\bb$, and which are identified as hadrons, contribute to $\Gbb$:
\begin{eqnarray}
\label{eq:gbbhs}
\Gbb^{\mathrm{meas}} =&& \Gbb^{\mathrm{zf}}
+ \mathrm{BR(Higgs}\rightarrow\bb) \times \nonumber\\
&& \left[
\mathrm{F}^{had}_{\ell\ell}(\MH) \Gll^{\mathrm{hs}}
+ \mathrm{F}^{had}_{\mathrm{inv}}(\MH) \Ginv^{\mathrm{hs}} \right] \,.
\end{eqnarray}
The result is shown in Figure~\ref{Mt_vs_Mh_hs6}.

The resulting dramatic upturn in the $\Gbb$ measurement band flattens the
lower extremity of the 68\% CL contour for the $\Zpole$ fit in which the
$\Afbzb$ measurement is dropped, but only modestly.
The lower $1\sigma$ error on $\Mt$ for this fit is 11\% smaller than it was when
Higgsstrahlung was ignored.
The corresponding shift in the lower $1\sigma$ error on $\Mt$ for the base fit
including all $\Zpole$ measurements is only 1.7\%.
Other aspects of the error contours remain numerically unaffected.


\section{Conclusions}

Our study of the manner in which the $\Zpole$ measurements constrain the
Standard Model parameters $\Mt$ and $\MH$ through their effect on radiative
corrections shows that the measured leptonic width, $\Gll$ plays an important,
 and non-linear role.
This non-linearity at low values of $\MH$ contributes to a surprising
stability in the agreement between the direct and indirect measurements of
$\Mt$ under variations in $\swsqeffl$.
This singular demonstration of the consistency of our experimental and
theoretical understanding of electroweak radiative corrections therefore
remains undisturbed by the apparent discrepancy between the values of
$\swsqeffl$ derived from the measurements of $\Afbzb$ and $\ALR$.

The neglect of external Higgstrahlung in ZFITTER and TOPAZ0 in the analysis
of the $\Zpole$ measurements is found to have negligible impact on the central
values of all the extracted Standard Model parameters, with the exception of
$\alfas$.
Within a CL of 68\% the error contours of these parameters also remain
essentially unaffected.

\vfill

\section*{Acknowledgments}

We would like to thank F. Jegerlehner, P. Gambino, and members of
the LEP Electroweak Working Group for useful discussions.
We also wish to thank the OPAL Collaboration, of which we are
both members, for access to detector-level Monte-Carlo simulations
of the Higgsstrahlung process.
In addition to the support staff at our own institutions we are pleased
to acknowledge the Department of Energy, USA, the Japanese Ministry of
Education, Culture, Sports, Science and
Technology (MEXT) and a grant under the MEXT International
Science Research Program,
Japanese Society for the Promotion of Science (JSPS).

\vfill

\bibliography{physrep}

\begin{thebibliography}{28}
\expandafter\ifx\csname natexlab\endcsname\relax\def\natexlab#1{#1}\fi
\expandafter\ifx\csname bibnamefont\endcsname\relax
  \def\bibnamefont#1{#1}\fi
\expandafter\ifx\csname bibfnamefont\endcsname\relax
  \def\bibfnamefont#1{#1}\fi
\expandafter\ifx\csname citenamefont\endcsname\relax
  \def\citenamefont#1{#1}\fi
\expandafter\ifx\csname url\endcsname\relax
  \def\url#1{\texttt{#1}}\fi
\expandafter\ifx\csname urlprefix\endcsname\relax\def\urlprefix{URL }\fi
\providecommand{\bibinfo}[2]{#2}
\providecommand{\eprint}[2][]{\url{#2}}

\bibitem[{\citenamefont{Ross and Veltman}(1975)}]{Ross:1975fq}
\bibinfo{author}{\bibfnamefont{D.~A.} \bibnamefont{Ross}} \bibnamefont{and}
  \bibinfo{author}{\bibfnamefont{M.~J.~G.} \bibnamefont{Veltman}},
  \bibinfo{journal}{Nucl. Phys.} \textbf{\bibinfo{volume}{B95}},
  \bibinfo{pages}{135} (\bibinfo{year}{1975}).

\bibitem[{LEP(2003)}]{LEPSMHIGGS}
\bibinfo{journal}{Phys. Lett.} \textbf{\bibinfo{volume}{B565}},
  \bibinfo{pages}{61} (\bibinfo{year}{2003}), \eprint{hep-ex/0306033}.

\bibitem[{\citenamefont{Bardin et~al.}(1989)\citenamefont{Bardin, Bilenkii,
  Mitselmakher, Riemann, and Sachwitz}}]{Bardin:1989di}
\bibinfo{author}{\bibfnamefont{D.~Y.} \bibnamefont{Bardin}},
  \bibinfo{author}{\bibfnamefont{M.~S.} \bibnamefont{Bilenkii}},
  \bibinfo{author}{\bibfnamefont{G.}~\bibnamefont{Mitselmakher}},
  \bibinfo{author}{\bibfnamefont{T.}~\bibnamefont{Riemann}}, \bibnamefont{and}
  \bibinfo{author}{\bibfnamefont{M.}~\bibnamefont{Sachwitz}},
  \bibinfo{journal}{Z. Phys.} \textbf{\bibinfo{volume}{C44}},
  \bibinfo{pages}{493} (\bibinfo{year}{1989}).

\bibitem[{\citenamefont{Bardin et~al.}(1990)\citenamefont{Bardin, Bilenkii,
  Riemann, Sachwitz, and Vogt}}]{Bardin:1990tq}
\bibinfo{author}{\bibfnamefont{D.~Y.} \bibnamefont{Bardin}},
  \bibinfo{author}{\bibfnamefont{M.~S.} \bibnamefont{Bilenkii}},
  \bibinfo{author}{\bibfnamefont{T.}~\bibnamefont{Riemann}},
  \bibinfo{author}{\bibfnamefont{M.}~\bibnamefont{Sachwitz}}, \bibnamefont{and}
  \bibinfo{author}{\bibfnamefont{H.}~\bibnamefont{Vogt}},
  \bibinfo{journal}{Comput. Phys. Commun.} \textbf{\bibinfo{volume}{59}},
  \bibinfo{pages}{303} (\bibinfo{year}{1990}).

\bibitem[{\citenamefont{Bardin et~al.}(1991{\natexlab{a}})}]{Bardin:1991fu}
\bibinfo{author}{\bibfnamefont{D.~Y.} \bibnamefont{Bardin}}
  \bibnamefont{et~al.}, \bibinfo{journal}{Nucl. Phys.}
  \textbf{\bibinfo{volume}{B351}}, \bibinfo{pages}{1}
  (\bibinfo{year}{1991}{\natexlab{a}}), \eprint{arXiv:hep-ph/9801208}.

\bibitem[{\citenamefont{Bardin et~al.}(1991{\natexlab{b}})}]{Bardin:1991de}
\bibinfo{author}{\bibfnamefont{D.~Y.} \bibnamefont{Bardin}}
  \bibnamefont{et~al.}, \bibinfo{journal}{Phys. Lett.}
  \textbf{\bibinfo{volume}{B255}}, \bibinfo{pages}{290}
  (\bibinfo{year}{1991}{\natexlab{b}}), \eprint{arXiv:hep-ph/9801209}.

\bibitem[{\citenamefont{Bardin et~al.}(1992)}]{Bardin:1992jc}
\bibinfo{author}{\bibfnamefont{D.~Y.} \bibnamefont{Bardin}}
  \bibnamefont{et~al.}, \emph{\bibinfo{title}{Zfitter: An analytical program
  for fermion pair production in $\ee$ annihilation}} (\bibinfo{year}{1992}),
  \eprint{arXiv:hep-ph/9412201}.

\bibitem[{\citenamefont{Bardin et~al.}(2001)}]{Bardin:1999yd}
\bibinfo{author}{\bibfnamefont{D.~Y.} \bibnamefont{Bardin}}
  \bibnamefont{et~al.}, \bibinfo{journal}{Comput. Phys. Commun.}
  \textbf{\bibinfo{volume}{133}}, \bibinfo{pages}{229} (\bibinfo{year}{2001}),
  \bibinfo{note}{recently updated with results from~\cite{Arbuzov}},
  \eprint{arXiv:hep-ph/9908433}.

\bibitem[{\citenamefont{Montagna
  et~al.}(1993{\natexlab{a}})\citenamefont{Montagna, Piccinini, Nicrosini,
  Passarino, and Pittau}}]{Montagna:1993py}
\bibinfo{author}{\bibfnamefont{G.}~\bibnamefont{Montagna}},
  \bibinfo{author}{\bibfnamefont{F.}~\bibnamefont{Piccinini}},
  \bibinfo{author}{\bibfnamefont{O.}~\bibnamefont{Nicrosini}},
  \bibinfo{author}{\bibfnamefont{G.}~\bibnamefont{Passarino}},
  \bibnamefont{and} \bibinfo{author}{\bibfnamefont{R.}~\bibnamefont{Pittau}},
  \bibinfo{journal}{Nucl. Phys.} \textbf{\bibinfo{volume}{B401}},
  \bibinfo{pages}{3} (\bibinfo{year}{1993}{\natexlab{a}}).

\bibitem[{\citenamefont{Montagna
  et~al.}(1993{\natexlab{b}})\citenamefont{Montagna, Piccinini, Nicrosini,
  Passarino, and Pittau}}]{Montagna:1993ai}
\bibinfo{author}{\bibfnamefont{G.}~\bibnamefont{Montagna}},
  \bibinfo{author}{\bibfnamefont{F.}~\bibnamefont{Piccinini}},
  \bibinfo{author}{\bibfnamefont{O.}~\bibnamefont{Nicrosini}},
  \bibinfo{author}{\bibfnamefont{G.}~\bibnamefont{Passarino}},
  \bibnamefont{and} \bibinfo{author}{\bibfnamefont{R.}~\bibnamefont{Pittau}},
  \bibinfo{journal}{Comput. Phys. Commun.} \textbf{\bibinfo{volume}{76}},
  \bibinfo{pages}{328} (\bibinfo{year}{1993}{\natexlab{b}}).

\bibitem[{\citenamefont{Montagna et~al.}(1996)\citenamefont{Montagna,
  Nicrosini, Passarino, and Piccinini}}]{Montagna:1996ja}
\bibinfo{author}{\bibfnamefont{G.}~\bibnamefont{Montagna}},
  \bibinfo{author}{\bibfnamefont{O.}~\bibnamefont{Nicrosini}},
  \bibinfo{author}{\bibfnamefont{G.}~\bibnamefont{Passarino}},
  \bibnamefont{and}
  \bibinfo{author}{\bibfnamefont{F.}~\bibnamefont{Piccinini}},
  \bibinfo{journal}{Comput. Phys. Commun.} \textbf{\bibinfo{volume}{93}},
  \bibinfo{pages}{120} (\bibinfo{year}{1996}), \eprint{arXiv:hep-ph/9506329}.

\bibitem[{\citenamefont{Montagna et~al.}(1999)\citenamefont{Montagna,
  Nicrosini, Piccinini, and Passarino}}]{Montagna:1998kp}
\bibinfo{author}{\bibfnamefont{G.}~\bibnamefont{Montagna}},
  \bibinfo{author}{\bibfnamefont{O.}~\bibnamefont{Nicrosini}},
  \bibinfo{author}{\bibfnamefont{F.}~\bibnamefont{Piccinini}},
  \bibnamefont{and}
  \bibinfo{author}{\bibfnamefont{G.}~\bibnamefont{Passarino}},
  \bibinfo{journal}{Comput. Phys. Commun.} \textbf{\bibinfo{volume}{117}},
  \bibinfo{pages}{278} (\bibinfo{year}{1999}), \bibinfo{note}{recently updated
  to include initial state pair radiation (G. Passarino, priv. comm.)},
  \eprint{arXiv:hep-ph/9804211}.

\bibitem[{\citenamefont{Burkhardt and Pietrzyk}(2001)}]{bib-BP01}
\bibinfo{author}{\bibfnamefont{H.}~\bibnamefont{Burkhardt}} \bibnamefont{and}
  \bibinfo{author}{\bibfnamefont{B.}~\bibnamefont{Pietrzyk}},
  \bibinfo{journal}{Phys. Lett.} \textbf{\bibinfo{volume}{B513}},
  \bibinfo{pages}{46} (\bibinfo{year}{2001}).

\bibitem[{\citenamefont{Veltman}(1977)}]{Veltman:1977kh}
\bibinfo{author}{\bibfnamefont{M.~J.~G.} \bibnamefont{Veltman}},
  \bibinfo{journal}{Nucl. Phys.} \textbf{\bibinfo{volume}{B123}},
  \bibinfo{pages}{89} (\bibinfo{year}{1977}).

\bibitem[{\citenamefont{Abbiendi et~al.}(2001)}]{Abbiendi:2000hu}
\bibinfo{author}{\bibfnamefont{G.}~\bibnamefont{Abbiendi}} \bibnamefont{et~al.}
  (\bibinfo{collaboration}{OPAL}), \bibinfo{journal}{Eur. Phys. J.}
  \textbf{\bibinfo{volume}{C19}}, \bibinfo{pages}{587} (\bibinfo{year}{2001}),
  \eprint{arXiv:hep-ex/0012018}.

\bibitem[{\citenamefont{Burgers and Jegerlehner}(1989)}]{Burgers:LEP1YR89VOL1}
\bibinfo{author}{\bibfnamefont{G.}~\bibnamefont{Burgers}} \bibnamefont{and}
  \bibinfo{author}{\bibfnamefont{F.}~\bibnamefont{Jegerlehner}}, in
  \emph{\bibinfo{booktitle}{Z PHYSICS AT LEP 1. PROCEEDINGS, WORKSHOP, GENEVA,
  SWITZERLAND, SEPTEMBER 4-5, 1989. VOL. 1: STANDARD PHYSICS}}, edited by
  \bibinfo{editor}{\bibfnamefont{G.}~\bibnamefont{Altarelli}},
  \bibinfo{editor}{\bibfnamefont{R.}~\bibnamefont{Kleiss}}, \bibnamefont{and}
  \bibinfo{editor}{\bibfnamefont{C.}~\bibnamefont{Verzegnassi}}
  (\bibinfo{publisher}{CERN}, \bibinfo{address}{Geneva, Switzerland},
  \bibinfo{year}{1989}), p.~\bibinfo{pages}{55}, \bibinfo{note}{yellow Report
  CERN 89-08}.

\bibitem[{\citenamefont{Jegerlehner}(1991{\natexlab{a}})}]{Jegerlehner:1991ed}
\bibinfo{author}{\bibfnamefont{F.}~\bibnamefont{Jegerlehner}},
  \bibinfo{journal}{Prog. Part. Nucl. Phys.} \textbf{\bibinfo{volume}{27}},
  \bibinfo{pages}{1} (\bibinfo{year}{1991}{\natexlab{a}}).

\bibitem[{\citenamefont{Jegerlehner}(1991{\natexlab{b}})}]{Jegerlehner:1991dq}
\bibinfo{author}{\bibfnamefont{F.}~\bibnamefont{Jegerlehner}}, in
  \emph{\bibinfo{booktitle}{Testing the Standard Model - TASI-90, proceedings:
  Theoretical Advanced Study Institute in Elementary Particle Physics, Boulder,
  Colo., Jun 3-27, 1990}}, edited by
  \bibinfo{editor}{\bibfnamefont{M.}~\bibnamefont{Cvetic}} \bibnamefont{and}
  \bibinfo{editor}{\bibfnamefont{P.}~\bibnamefont{Langacker}}
  (\bibinfo{publisher}{World Scientific}, \bibinfo{address}{Singapore},
  \bibinfo{year}{1991}{\natexlab{b}}), p. \bibinfo{pages}{916},
  \bibinfo{note}{pSI-PR-91-08}.

\bibitem[{\citenamefont{Bardin et~al.}(1999)\citenamefont{Bardin,
  Gr{\"u}newald, and Passarino}}]{PCP99}
\bibinfo{author}{\bibfnamefont{D.~Y.} \bibnamefont{Bardin}},
  \bibinfo{author}{\bibfnamefont{M.}~\bibnamefont{Gr{\"u}newald}},
  \bibnamefont{and}
  \bibinfo{author}{\bibfnamefont{G.}~\bibnamefont{Passarino}},
  \emph{\bibinfo{title}{Precision calculation project report}}
  (\bibinfo{year}{1999}), \eprint{arXiv:hep-ph/9902452}.

\bibitem[{\citenamefont{Chetyrkin et~al.}(1995)}]{bib-PCLI-QCD}
\bibinfo{author}{\bibfnamefont{K.}~\bibnamefont{Chetyrkin}}
  \bibnamefont{et~al.}, in \emph{\bibinfo{booktitle}{Reports of the working
  group on precision calculations for the Z resonance}}, edited by
  \bibinfo{editor}{\bibfnamefont{D.}~\bibnamefont{Bardin}},
  \bibinfo{editor}{\bibfnamefont{W.}~\bibnamefont{Hollik}}, \bibnamefont{and}
  \bibinfo{editor}{\bibfnamefont{G.}~\bibnamefont{Passarino}}
  (\bibinfo{publisher}{CERN}, \bibinfo{address}{Geneva, Switzerland},
  \bibinfo{year}{1995}), \bibinfo{number}{CERN 95-03}, p. \bibinfo{pages}{175},
  \bibinfo{note}{yellow Report CERN 95-03}.

\bibitem[{Gro()}]{Group:2003ih}
\bibinfo{note}{The LEP Collaborations ALEPH, DELPHI, L3, OPAL and the LEP
  Electroweak Working Group, and the SLD Heavy Flavour and Electroweak Groups,
  {\it A Combination of Preliminary Electroweak Measurements and Constraints on
  the Standard Model}, hep-ex/0312023}.

\bibitem[{\citenamefont{Abe et~al.}(2001)}]{ref:sld-al2000}
\bibinfo{author}{\bibfnamefont{K.}~\bibnamefont{Abe}} \bibnamefont{et~al.}
  (\bibinfo{collaboration}{SLD}), \bibinfo{journal}{Phys. Rev. Lett.}
  \textbf{\bibinfo{volume}{86}}, \bibinfo{pages}{1162} (\bibinfo{year}{2001}),
  \eprint{arXiv:hep-ex/0010015}.

\bibitem[{\citenamefont{Hagiwara et~al.}(2002)}]{PDG2002}
\bibinfo{author}{\bibfnamefont{K.}~\bibnamefont{Hagiwara}} \bibnamefont{et~al.}
  (\bibinfo{collaboration}{Particle Data Group}), \bibinfo{journal}{Phys. Rev.}
  \textbf{\bibinfo{volume}{D66}}, \bibinfo{pages}{010001}
  (\bibinfo{year}{2002}).

\bibitem[{PP-()}]{PP-MW-GW:combination}
\bibinfo{note}{Combination of CDF snd D\O\ Results on W Boson Mass and Width,
  Tevatron Electroweak Working Group and the CDF and D\O\ Collaborations, CDF
  Note 5888, D\O\ Note 3963, FERMILAB-FN-0716, July 2002}.

\bibitem[{\citenamefont{Franzini and Taxil}(1989)}]{LEP1YR89VOL2}
\bibinfo{author}{\bibfnamefont{P.}~\bibnamefont{Franzini}} \bibnamefont{and}
  \bibinfo{author}{\bibfnamefont{P.}~\bibnamefont{Taxil}}, in
  \emph{\bibinfo{booktitle}{Z PHYSICS AT LEP 1. PROCEEDINGS, WORKSHOP, GENEVA,
  SWITZERLAND, SEPTEMBER 4-5, 1989. VOL. 2: HIGGS SEARCH AND NEW PHYSICS}},
  edited by \bibinfo{editor}{\bibfnamefont{G.}~\bibnamefont{Altarelli}},
  \bibinfo{editor}{\bibfnamefont{R.}~\bibnamefont{Kleiss}}, \bibnamefont{and}
  \bibinfo{editor}{\bibfnamefont{C.}~\bibnamefont{Verzegnassi}}
  (\bibinfo{publisher}{CERN}, \bibinfo{address}{Geneva, Switzerland},
  \bibinfo{year}{1989}), p.~\bibinfo{pages}{84}, \bibinfo{note}{yellow Report
  CERN 89-08}.

\bibitem[{\citenamefont{Ahmet et~al.}(1991)}]{Ahmet:1991eg}
\bibinfo{author}{\bibfnamefont{K.}~\bibnamefont{Ahmet}} \bibnamefont{et~al.}
  (\bibinfo{collaboration}{OPAL}), \bibinfo{journal}{Nucl. Instrum. Meth.}
  \textbf{\bibinfo{volume}{A305}}, \bibinfo{pages}{275} (\bibinfo{year}{1991}).

\bibitem[{\citenamefont{Janot}(1989)}]{hzha}
\bibinfo{author}{\bibfnamefont{P.}~\bibnamefont{Janot}}, in
  \emph{\bibinfo{booktitle}{Physics at LEP, CERN 96-01, vol. 2}}, edited by
  \bibinfo{editor}{\bibfnamefont{T.~S.} \bibnamefont{G.~Altarelli}}
  \bibnamefont{and} \bibinfo{editor}{\bibfnamefont{F.}~\bibnamefont{Zwirner}}
  (\bibinfo{publisher}{CERN}, \bibinfo{address}{Geneva, Switzerland},
  \bibinfo{year}{1989}), p. \bibinfo{pages}{309}, \bibinfo{note}{for a
  description of the updates and for the code see:\\
  http://alephwww.cern.ch/$\sim$janot/Generators.html}.

\bibitem[{\citenamefont{Arbuzov}(1999)}]{Arbuzov}
\bibinfo{author}{\bibfnamefont{A.~B.} \bibnamefont{Arbuzov}},
  \emph{\bibinfo{title}{Light pair corrections to electron positron
  annihilation at lep/slc}} (\bibinfo{year}{1999}),
  \eprint{arXiv:hep-ph/9907500}.

\end{thebibliography}

\end{document}